\documentclass[journal=jacsat,manuscript=article]{achemso}

\usepackage{chemformula} 
\usepackage[T1]{fontenc} 
\usepackage{amssymb}
\usepackage{float}
\usepackage[T1]{fontenc} 
\usepackage[section]{placeins}

\DeclareUnicodeCharacter{2212}{\ensuremath{-}}

\title{When is the Four-phonon Effect in Half-Heusler Materials more Pronounced?}

\author{Yu Wu}
\email{wuyu9573@qq.com}
\affiliation{Yangtze Delta Region Institute (Huzhou), University of Electronic Science and Technology, Huzhou, Zhejiang 313001, China}

\author{Shengnan Dai}
\email{musenc@shu.edu.cn}
\affiliation{Materials Genome Institute, Shanghai University, Shanghai, 200444, China}

\author{Linxuan Ji}
\affiliation{School of Physics and State Key Laboratory of Electronic Thin Films and Integrated Devices, University of Electronic Science and Technology, Sichuan, Chengdu 610054, China}

\author{Yimin Ding}
\affiliation{Yangtze Delta Region Institute (Huzhou), University of Electronic Science and Technology, Huzhou, Zhejiang 313001, China}

\author{Jiong Yang}
\affiliation{Materials Genome Institute, Shanghai University, Shanghai, 200444, China}
\alsoaffiliation{Zhejiang Laboratory, Hangzhou, Zhejiang, 311100, China}

\author{Liujiang Zhou}
\email{ljzhou@uestc.edu.cn}
\affiliation{School of Physics and State Key Laboratory of Electronic Thin Films and Integrated Devices, University of Electronic Science and Technology, Sichuan, Chengdu 610054, China}
\alsoaffiliation{Yangtze Delta Region Institute (Huzhou), University of Electronic Science and Technology, Huzhou, Zhejiang 313001, China}

\begin{document}

\begin{abstract}

Suppressed three-phonon scattering processes have been considered to be the direct cause of materials exhibiting significant higher-order four-phonon interactions. However, after calculating the phonon-phonon interactions of 128 Half-Heusler materials by high-throughput, we find that the acoustic phonon bandwidth dominates the three-phonon and four-phonon scattering channels and keeps them roughly in a co-increasing or decreasing behavior. The $aao$ and $aaa$ three-phonon scattering channels in Half-Heusler materials are weakly affected by the acoustic-optical gap and acoustic bunched features respectively only when acoustic phonon bandwidths are close. Finally, we found that Half-Heusler materials with smaller acoustic bandwidths tend to have a more pronounced four-phonon effect, although three-phonon scattering may not be significantly suppressed at this time.

\end{abstract}

\flushbottom
\maketitle

\thispagestyle{empty}

\section*{Introduction}

Half-Heusler (HH) alloys are among the most promising thermoelectric (TE) materials for medium and high-temperature waste heat recovery applications\cite{Zhu2023,Dong2022,Li2020a,Xing2019}. The TE performance of a material is quantified by a dimensionless figure of merit, $ZT=S^2\sigma T/\kappa$, where $S$ represents the Seebeck coefficient, $\sigma$ is the electron conductivity, and $\kappa$ is the thermal conductivity contributed by carriers ($\kappa_e$) and lattice ($\kappa_L$). To achieve a high $ZT$ value, a large power factor ($S^2\sigma$) and low $\kappa$ should be achieved simultaneously in crystalline solids, which behave as ``phonon-glass and electron-crystal''\cite{Pei2011,Liu2012}. However, the tight coupling among the thermoelectric parameters $S$, $\sigma$, and $\kappa$ hinders the improvement of TE performance. HH materials exhibit multi-valley band structures because of their highly symmetric crystals\cite{Kumarasinghe2019,Brod2022}, which brings large $S$. Due to the suppressed electron-acoustic phonon coupling in HH materials, electron scattering from phonons is much weaker with the intrinsic low deformation potential\cite{Zhou2018,Zhu2023}. Therefore, decoupling of $S$ and $\sigma$ can be achieved in HH materials, leading to a large power factor. Although, the high lattice thermal transport of HH materials often limits their thermoelectric conversion efficiency\cite{Li2024a,Ren2022,Carrete2014}.

Understanding the lattice thermal transport mechanism in HH materials and high-throughput screening of HH materials with intrinsically low $\kappa_L$ is important for thermal management in HH devices. The weakly bound atoms in HH materials can be regarded as rattlers. The rattlers can reduce the phonon group velocities and introduce large lattice anharmonicity. Zhang et al. used two simple metrics containing atomic mass and phonon frequency information to represent these two mechanisms and established their connection to $\kappa_L$ in HH materials\cite{Feng2020}. Wang et al. revealed the softening effect of occupied d-d antibonding states on the acoustic branches of 19-electron HH materials\cite{Wang2024}. Miyazaki et al. developed a machine learning (ML) algorithm to predict the lattice parameters and $\kappa_L$ of HH materials from the atomic information of their constituent atoms\cite{Miyazaki2021}. However, few studies apply high-order four-phonon (4ph) interactions to HH materials on the calculations of $\kappa_L$. The rattlers in HH materials can induce avoided crossing behavior between the acoustic and optical branches, and the resulting acoustic-optical ($a-o$) gap has long been thought to be one of the triggers for the pronounced 4ph effect\cite{Feng2017,Ravichandran2020,Wu2023}. By studying the $\kappa_L$ of the 17 compounds with zinc blende structure, Broido et al. found that the 4ph effect is pronounced in the materials where three-phonon (3ph) scattering is suppressed due to the selection rule, coming from large $a-o$ gap, bunched features of acoustic branches and other factors\cite{Ravichandran2020}. However, the numerical relationship between these factors and the 4ph effect is not given due to the limited data. Recently, Hong et al. found in the HH material LuNiBi that flat parallel acoustic phonon branches can effectively suppress the 3ph scattering processes and highlight the 4ph effect\cite{Li2024}. However, because this phenomenon occurs in the high-frequency acoustic phonon region, the effect of 4ph scattering on the $\kappa_L$ is almost negligible. In any case, previous studies have argued that the pronounced 4ph effect is always accompanied by suppressed 3ph scattering processes. Confusingly, Xia et al. controlled the $a-o$ gap of HgTe by multiplying the longitudinal acoustic branch by different constants and found that a large $a-o$ gap did not make the 4ph effect more significant, and even had a negative impact\cite{Xia2020}. Several materials are reported in the literature with significant 4ph effects that do not exhibit the features of suppressed 3ph scattering from the phonon dispersion\cite{Yue2023,Zhou2022,Yang2024}.

Here, we study the lattice thermal properties of 128 stable HH materials in the MatHub-3d database\cite{Yao2021}. The results indicate that the phonon bandwidth controls the 3ph and 4ph scattering channels, and generally causes them to exhibit simultaneous increase or decrease behavior rather than a mutually exclusive growth. The suppression of the 3ph scattering channels by the large $a-o$ gap and the bunched features is evident only when the acoustic bandwidth is close. Ultimately, we find that the 4ph effect tends to be more pronounced in HH materials with smaller acoustic bandwidth, even though they also have relatively large 3ph scattering.

\section*{Results and Discussion}

\begin{figure*}[ht!]
\centering
\includegraphics[width=1\linewidth]{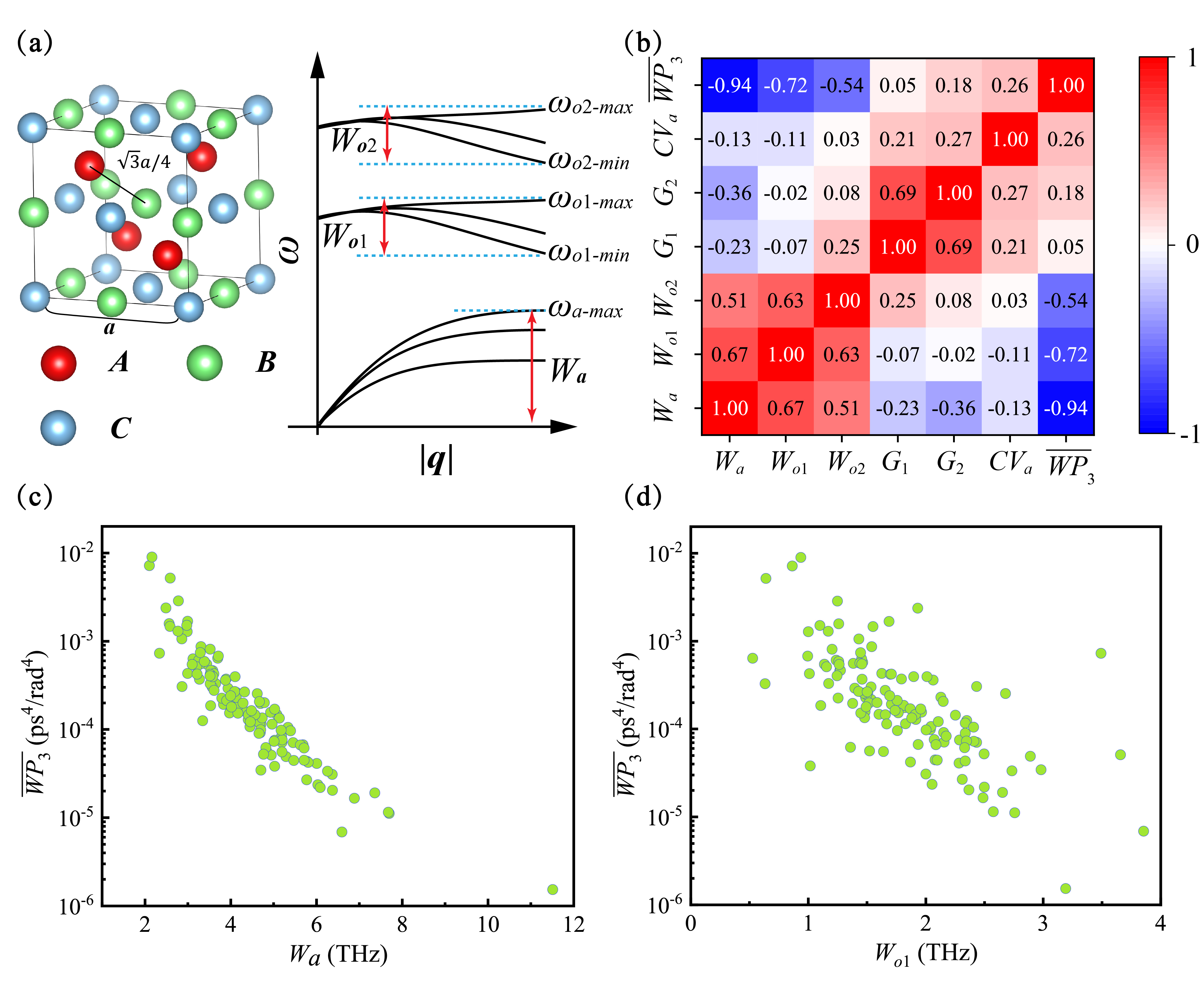}
\caption{(a) Left: The conventional cell of HH materials. Right: The schematic diagram of phonon dispersions of HH materials. (b) The correlation matrix for different physical quantities reflecting the features of the phonon dispersions and the average weighted 3ph phase space of acoustic phonons $\overline{WP_3}$. (c) The $\overline{WP_3}$ plotted as a function of the bandwidth of acoustic branches $W_a$. (d) The $\overline{WP_3}$ plotted as a function of the bandwidth of the first group of optical branches $W_{o1}$.}
\label{Fig1}
\end{figure*}

HH materials are ternary intermetallics with a general $ABC$ formula and exhibit a cubic crystal structure belonging to the $F\overline{4}3m$ space group as shown in Fig.~\ref{Fig1}(a). The structure consists of three interpenetrating face-centered cubic (fcc) sublattices, as well as an additional unoccupied fcc sublattice. The $B$ and $C$ atoms are equivalent, while the $A$ atom is not equivalent to them. Exchanging the $A$ atom with either $B$ or $C$ will result in a different material. The nearest $B-C$ bond is a/2 while the nearest $A-B$ and $A-C$ bonds are $\sqrt{3}a/4$. Since there are 3 atoms in the primitive cell of HH materials, the phonon dispersion consists of 9 phonon branches, as illustrated in the schematic diagram on the right side of Fig.~\ref{Fig1}(a). Several physical quantities are constructed to reflect phonon spectral features. These include the phonon bandwidth of acoustic branches $W_a=\omega_{a-max}$, the phonon bandwidth of the first ($W_{o1}=\omega_{o1-max}-\omega_{o1-min}$) and the second group of optical branches ($W_{o2}=\omega_{o2-max}-\omega_{o2-min}$), the $G_1=\omega_{o1-min}/\omega_{a-max}$ and $G_2=\omega_{o2-min}/\omega_{a-max}$ reflecting the relative magnitude of the $a-o$ bandgap, and the coefficient of variation of acoustic phonon frequency $CV_a=\frac{1}{N_{\boldsymbol{q}}} \sum_{\boldsymbol{q}} \frac{\sigma\left(\omega_{n, \boldsymbol{q}}\right)}{\mu\left(\omega_{n, \boldsymbol{q}}\right)}\;(n=1,2,3)$ reflecting the bunched features, where $\omega_{a-max}$ is the maximum frequency of  acoustic phonons, $\omega_{o1-max}$ ($\omega_{o1-min}$) and $\omega_{o2-max}$ ($\omega_{o2-min}$) are the maximum (minimum) frequency of the first and second group of optical branches, $\omega_{n, \boldsymbol{q}}$ is the frequency of phonons with branch $n$ and wave vector $\boldsymbol{q}$, $N_{\boldsymbol{q}}$ is the number of $\boldsymbol{q}$, $\sigma$ and $\mu$ represent standard deviation and average, respectively. The correlation matrix for the above physical quantities and the average weighted 3ph phase space of acoustic phonons $\overline{WP_3}$ at 300 K is shown in Fig.~\ref{Fig1}(b). The $WP_3$ is definded in Ref.$\rm{\cite{Li2015a}}$ (Eq. S1 and S2). The $\overline{WP_3}$ exhibits a clear negative correlation with phonon bandwidths $W_a$ and $W_{o1}$ while a weak correlation with other parameters. Figure~\ref{Fig1}(c) and (d) show the $\overline{WP_3}$ as a function of the $W_a$ and $W_{o1}$, respectively. It indicates that small phonon bandwidths, especially $W_a$, contribute to obtaining larger acoustic phonon scattering channels in HH materials. Figure S1 shows the average 3ph phase space of acoustic phonons $\overline{P_3}$ as a function of $W_a$, which shows a similar trend to $\overline{WP_3}$. Compared with $WP_3$, $P_3$ ignores the influence of the phonon population and the absolute phonon frequency\cite{Li2015a,shengbte2014}, which further suggests that the $\delta$ function in the $WP_3$ plays an important role in the variation of $\overline{WP_3}$ with $W_a$. The $\delta$ function is related to the conservation of energy and momentum in the 3ph interaction process.

\begin{figure*}[ht!]
\centering
\includegraphics[width=1\linewidth]{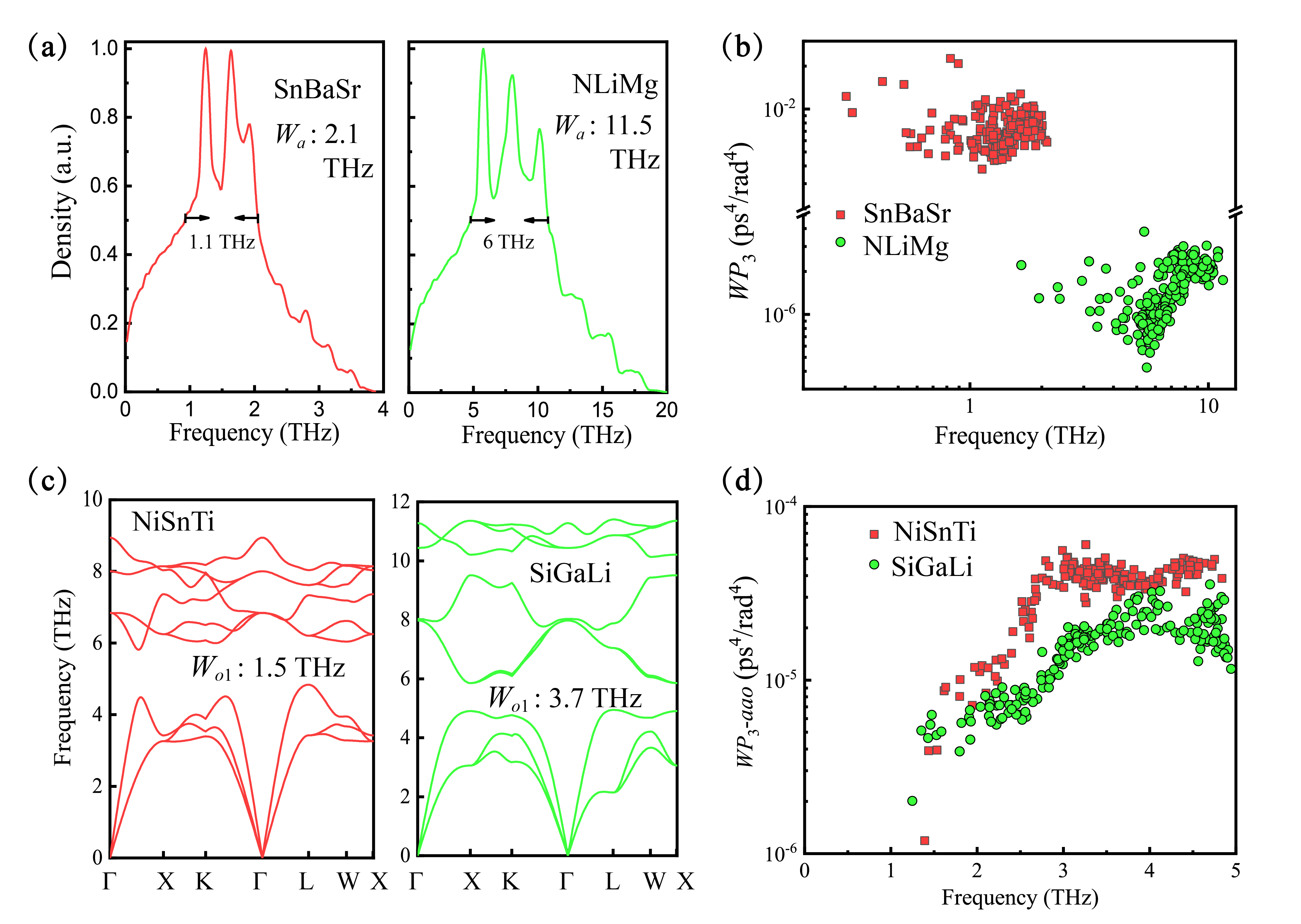}
\caption{(a) The frequency distribution contributed by two acoustic phonons in the $\Gamma$-K direction for SnBaSr and NLiMg. (b) The $WP_3$ of acoustic phonons for SnBaSr and NLiMg. (c) The phonon dispersion of NiSnTi and SiGaLi. (d) $WP_3$-$aao$ of acoustic phonons for NiSnTi and SiGaLi.}
\label{Fig2}
\end{figure*}

To understand the influence of $W_a$ on the $WP_3$, Fig.~\ref{Fig2}(a) shows the frequency distribution contributed by two acoustic phonons in the $\Gamma$-K direction for SnBaSr and NLiMg. The horizontal coordinate is the sum of the frequencies of the two phonons and the vertical coordinate is the normalization of the sum of the phonon population, which reflects the frequency density generated by two phonons. The SnBaSr with smaller $W_a$ has a more concentrated frequency distribution than NLiMg. This is favorable for the energy conservation conditions of the phonon scattering process, in other words, the third phonon falling in a more concentrated phonon distribution will have a better chance to participate in the 3ph interaction. As a result, the $WP_3$ of acoustic phonons in SnBaSr is nearly 4 orders of magnitude higher than that of NLiMg as seen in Fig.~\ref{Fig2}(b). The optical branch bandwidth mainly affects the $aao$ process in 3ph scattering. Small optical branch bandwidth is accompanied by flat spectral features, in which case the phonon frequency is insensitive to changes in the wavevector $\boldsymbol{q}$. In the absorption process where a mode ($\boldsymbol{q_0}, \omega_0$) is combined with ($\boldsymbol{q_1}, \omega_1$) to form ($\boldsymbol{q_2}, \omega_2$), for any ($\boldsymbol{q_1}, \omega_{flat}-\omega_0$), one can easily find a mode ($\boldsymbol{q_0}+\boldsymbol{q_1}, \omega_{flat}$) in the flat optical modes region that enables the scattering to occur. Taking NiSnTi and SiGaLi as an example, they have close $W_a$ but quite different $W_{o1}$ as seen in Fig.~\ref{Fig2}(c). In Fig.~\ref{Fig2}(d), the NiSnTi has a smaller $W_{o1}$ and therefore a larger scattering phase space of $aao$ process.

\begin{figure*}[ht!]
\centering
\includegraphics[width=1\linewidth]{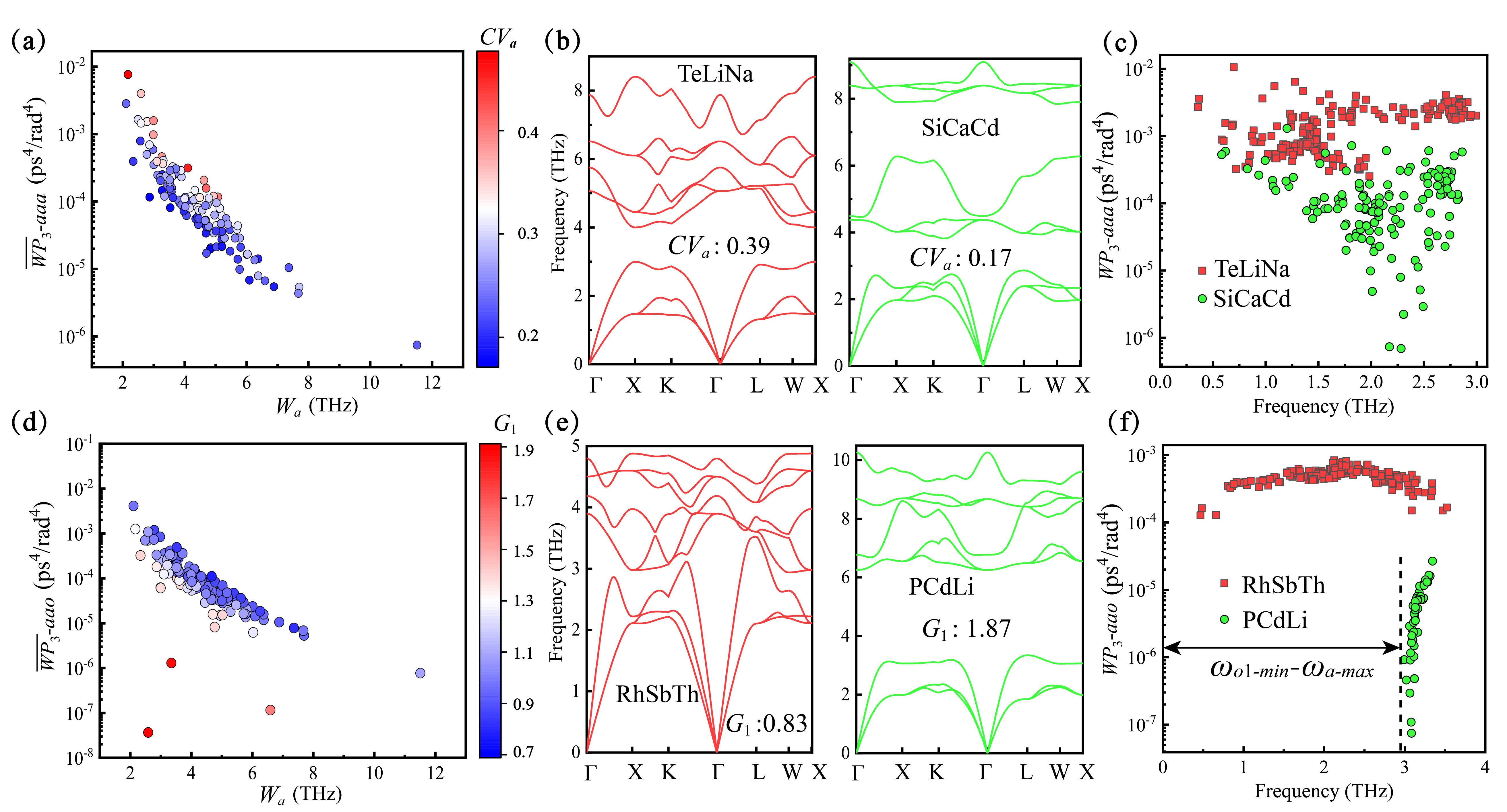}
\caption{(a) The $\overline{WP_3}$-$aaa$ plotted as a function of $W_a$. The color of the scatters represents the magnitude of the $CV_a$. (b) The phonon dispersion of TeLiNa and SiCaCd. (c) The $WP_3$-$aaa$ of acoustic phonons for TeLiNa and SiCaCd. (d) The $\overline{WP_3}$-$aao$ plotted as a function of $W_a$. The color of the scatters represents the magnitude of the $G_1$. (e) The phonon dispersion of RhSbTh and PCdLi. (f) The $WP_3$-$aao$ of acoustic phonons for RhSbTh and PCdLi.}
\label{Fig3}
\end{figure*}

The suppression of $aaa$ scattering processes by the bunched features of acoustic branches and the suppression of $aao$ scattering processes by large $a-o$ gaps have been considered the main reasons for limiting the 3ph interactions of materials\cite{Feng2017,Ravichandran2020,He2024}. Instead, we find that the phonon branch bandwidth dominates $WP_3$. To clarify the bunched features and the effect of $a-o$ gap on $WP_3$, Fig.~\ref{Fig3}(a) shows the distribution of $\overline{WP_3}$-$aaa$ with $W_a$ projected by the magnitude of $CV_a$. It can be seen that the magnitude of $CV_a$ only has a more pronounced effect on the $aaa$ scattering channel in HH materials when $W_a$ approaches. Figure~\ref{Fig3}(b) shows the phonon dispersions of TeLiNa and SiCaCd, which have similar $W_a$ but large differences in $CV_a$. The $WP_3$-$aaa$ of SiCaCd with smaller $CV_a$ is generally less than that of TeLiNa and shows a sharp drop near 2.2 THz. The bunched features effect on the $aaa$ scattering process is analogous to a straightforward argument that three phonons from the same acoustic branch cannot simultaneously satisfy both momentum and energy conservation\cite{Lax1981,Ravichandran2020,Ji2024}. Similarly, Fig.~\ref{Fig3}(d) shows the distribution of $\overline{WP_3}$-$aao$ with $W_a$ for the $G_1$ projection. Except for individual materials with particularly large $G_1$, the $G_1$ only has a more pronounced effect on the $aao$ scattering channels in HH materials when $W_a$ approaches. In addition, although the $\overline{WP_3}$-$aao$ of individual materials is ultralow due to the extremely high $G_1$ value, the total $\overline{WP_3}$ does not show a mutation with $W_a$ due to the relatively continuous change trend of $\overline{WP_3}$-$aaa$ with $W_a$. Figure~\ref{Fig3}(e) shows the phonon dispersions of RhSbTh and PCdLi. Compared with RhSbTh, PCdLi has an obvious open $a-o$ gap. The results of $WP_3$-$aao$ shown in Fig.~\ref{Fig3}(f) indicate that the acoustic phonons with frequency lower than the $a-o$ gap ($\omega_{o1-min}-\omega_{a-max}$) can not participate in the $aao$ process.

\begin{figure*}[ht!]
\centering
\includegraphics[width=1\linewidth]{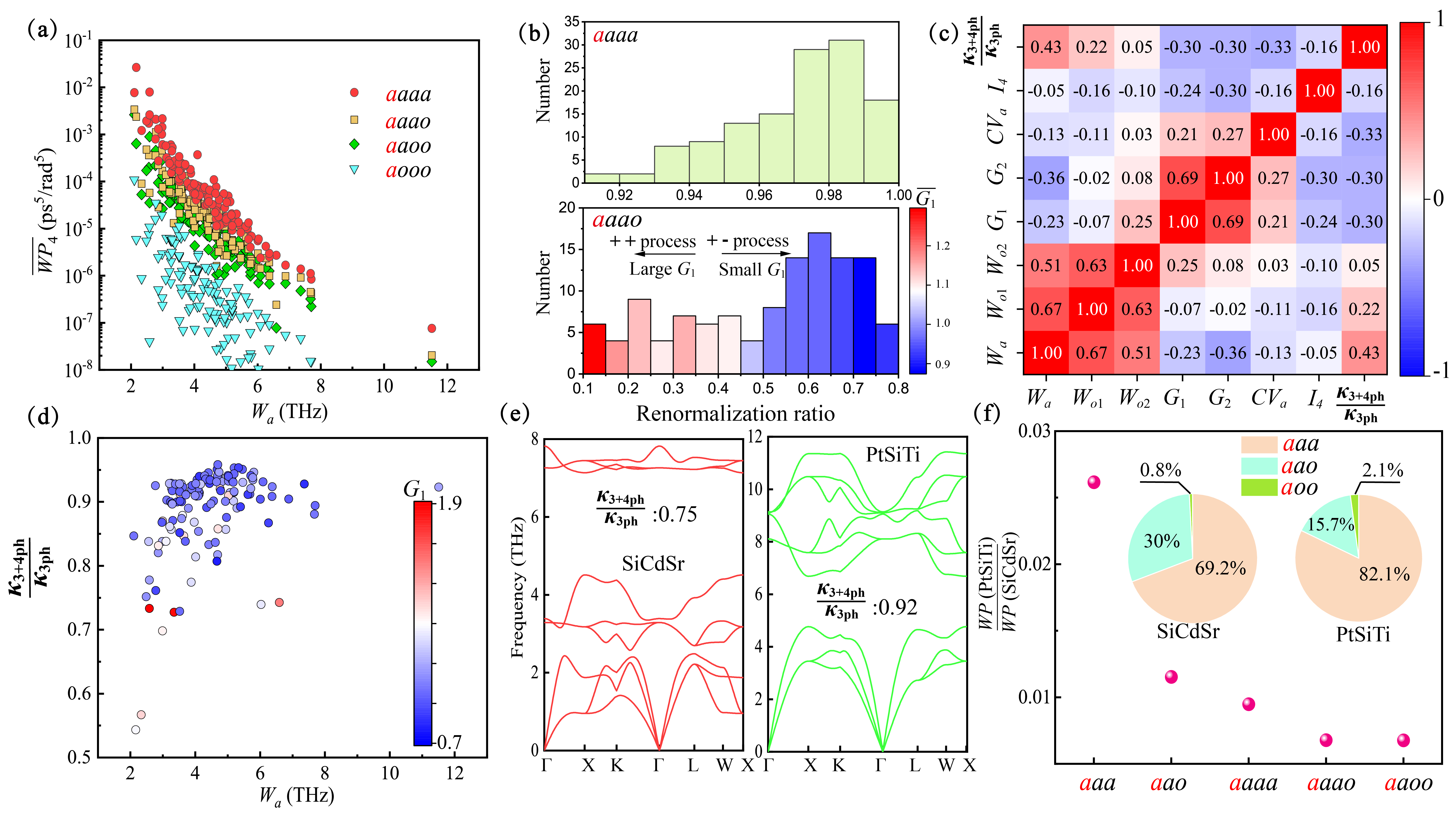}
\caption{(a) The channel-resolved average weighted 4ph phase space of acoustic phonons $\overline{WP_4}$ plotted as a function of $W_a$. (b) The statistical histogram of the renormalization ratio for $aaaa$ and $aaao$ process. (c) The correlation matrix for different physical quantities and the $\frac{\kappa_{3+4 \mathrm{ph}}}{\kappa_{3 \mathrm{ph}}}$. (d) The $\frac{\kappa_{3+4 \mathrm{ph}}}{\kappa_{3 \mathrm{ph}}}$ plotted as a function of $W_a$. The color of the scatters represents the magnitude of the $G_1$. (e) The phonon dispersion of SiCdSr and PtSiTi. (f) The ratio of weighted phase for different scattering processes of PtSiTi and SiCdSr. The inset shows the proportion of $aaa$, $aao$, $aoo$ in $\overline{WP_3}$ for acoustic phonons in SiCdSr and PtSiTi.
}
\label{Fig4}
\end{figure*}

The channel-resolved average weighted 4ph phase space of acoustic phonons $\overline{WP_4}$ plotted as a function of $W_a$ is shown in Fig.~\ref{Fig4}(a). The $aaaa$, $aaao$ and $aaoo$ process dominate $\overline{WP_4}$ and show a similar trend to $\overline{WP_3}$. The negative correlation between $\overline{WP_4}$ and $W_a$ is also because a narrow bandwidth makes it easier to realize the energy conservation condition during scattering. Therefore, $\overline{WP_3}$ and $\overline{WP_4}$ in HH materials roughly increase and decrease with the change of $W_a$. The large $a-o$ gap as well as the bunched features may not be a signal of a more pronounced 4ph effect. The statistical histograms of the renormalization ratio for $aaaa$ and $aaao$ process are shown in Fig.~\ref{Fig4}(b). The results for the $aaoo$ process are shown in Fig.~S2. The $aaaa$ and  $aaoo$ process occurs mainly through phonon renormalization (+ -). For the $aaao$ process, materials with large $G_1$ are usually accompanied by large phonon absorption (+ +), while small $G_1$ is usually accompanied by phonon renormalization dominance. To figure out when the 4ph effect is more pronounced, the correlation matrix for different physical quantities and the $\frac{\kappa_{3+4 \mathrm{ph}}}{\kappa_{3 \mathrm{ph}}}$ is shown in Fig.~\ref{Fig4}(c). $\kappa_{3 \mathrm{ph}}$ and $\kappa_{3+4 \mathrm{ph}}$ are the lattice thermal conductivity considering 3ph and 3+4ph interaction, respectively. In addition to the quantities related to the phonon spectrum, the above physical quantities also include the $I_4$, which reflects the strength of the 4ph effect, defined as $\left|\Phi_4 / \Phi_3\right|^2 /\left|\Phi_2\right|$, where $\Phi$ is the interaction force constant (IFC). The correlation coefficient between $\frac{\kappa_{3+4 \mathrm{ph}}}{\kappa_{3 \mathrm{ph}}}$ and $I_4$ is extremely small, indicating that the effect of scattering strength on the 4ph effect is negligible compared to the scattering channels. $\frac{\kappa_{3+4 \mathrm{ph}}}{\kappa_{3 \mathrm{ph}}}$ is mainly positively correlated with $W_a$ and weakly negatively correlated with $G_1$. 
Notably, $\frac{\kappa_{3+4 \mathrm{ph}}}{\kappa_{3 \mathrm{ph}}}$ does not show a positive correlation with $CV_a$, suggesting that the bunched features do not lead to a significant 4ph effect at all. Figure~S3(a) shows the $\overline{WP_4}$-$aaaa$ plotted as a function of $W_a$ with the $CV_a$ projection. Combined with Fig.~\ref{Fig3}(a), it can be found that the bunched features also significantly restrict the $aaaa$ 4ph scattering channels while reducing the $aaa$ 3ph scattering channels. As seen in Fig~S3(b) and Fig.~\ref{Fig3}(d), the $aaao$ 4ph scattering channels is not as sensitive as $aao$ with $G_1$.
Figure~\ref{Fig4}(d) shows the $\frac{\kappa_{3+4 \mathrm{ph}}}{\kappa_{3 \mathrm{ph}}}$ plotted as a function of $W_a$. The color of the scatters represents the magnitude of the $G_1$. This further suggests that small acoustic phonon bandwidths tend to be accompanied by a more pronounced 4ph effect. 
Below, we give examples of reasons why a large $G_1$ may not lead to significant 4ph effects. Figure~\ref{Fig4}(e) shows the phonon dispersion of SiCdSr and PtSiTi. PtSiTi has large $G_1$ but small $\frac{\kappa_{3+4 \mathrm{ph}}}{\kappa_{3 \mathrm{ph}}}$ compared to SiCdSr. The ratio of weighted phase for different scattering processes of PtSiTi and SiCdSr is shown in Fig.~\ref{Fig4}(f). It can be seen that the scattering channels of PtSiTi are not only weaker than that of SiCdSr in $aao$ process but also in other scattering processes including 4ph interaction due to the larger $W_a$ of PtSiTi. The combined result is that the 4ph effect in PtSiTi is not obvious. The inset shows that the larger $G_1$ in PtSiTi simply brings about a smaller proportion of $aao$ processes in the 3ph scattering channel, which does not imply a more prominent 4ph effect.

\section*{Conclusions}

In conclusion, we studied the phonon-phonon interactions of 128 HH materials. The acoustic bandwidth controls the three and 4ph scattering channels in HH materials and keeps them roughly in a co-increasing and decreasing behavior. The $a-o$ gap and acoustic bunched features only affect the proportion of $aao$ and $aaa$ scattering processes in the total 3ph scattering, and show a limited influence on the absolute magnitude of 3ph scattering when the bandwidths are close. Therefore, the more pronounced 4ph effect does not necessarily occur in materials that exhibit suppressed 3ph scattering by large $a-o$ gap or acoustic bunched features. The final results show that the 4ph effect is more pronounced in HH materials with smaller acoustic bandwidths, although the 3ph scattering is not significantly suppressed. Our study has important implications for understanding the 4ph effect and thermal management in HH thermoelectric materials.

\section*{Numerical methods}

The calculations are implemented using the Vienna Ab Initio simulation package (VASP) based on density functional theory (DFT)\cite{Kresse1996} with the projector augmented wave (PAW) method and Perdew− Burke−Ernzerhof (PBE) exchange−correlation functional\cite{Perdew1996}. The cut-off energy of the plane wave is set to 500 eV. The energy convergence value between two consecutive steps is set as $10^{-5}\;$eV when optimizing atomic positions and the maximum Hellmann-Feynman (HF) force acting on each atom is $10^{-3}\;$eV/\r{A}. The calculations of $\kappa_L$ and other relevant parameters such as phonon relaxation time are carried out by the ShengBTE software\cite{shengbte2014,Han2022} which operates based on the iterative scheme. The $\bf{q}$-mesh in the first irreducible Brillouin Zone is set to be 12$\times$12$\times$12. A maximum likelihood estimation method is used to accelerate the calculation of the 4ph scattering rate\cite{Guo2024}. The sample size is set as $4\times10^{5}$. The recently introduced on-the-fly Machine Learning Potential (FMLP) of VASP is used to accelerate AIMD simulation, which has been verified to be reasonable for use in calculations of thermal transport properties\cite{Cui2023}. The supercells of 4$\times$4$\times$4 containing 192 atoms are chosen. The simulation is run for 20 ps with a timestep of 1 fs. The second-order, third-order, and fourth-order IFCs are determined from the AIMD simulation by using the TDEP method\cite{Hellman2013}. The cutoff radius of third-order and fourth-order IFCs takes into account the interaction between the fifth and fourth nearest neighbor atoms respectively.

\section*{Conflicts of interest}
There are no conflicts to declare.

\section*{Acknowledgements}

This work is supported by the Natural Science Foundation of China (12304038), Huzhou Natural Science Foundation (2023YZ50), the Startup funds of Outstanding Talents of UESTC (A1098531023601205), National Youth Talents Plan of China (G05QNQR049).



\providecommand{\latin}[1]{#1}
\makeatletter
\providecommand{\doi}
  {\begingroup\let\do\@makeother\dospecials
  \catcode`\{=1 \catcode`\}=2 \doi@aux}
\providecommand{\doi@aux}[1]{\endgroup\texttt{#1}}
\makeatother
\providecommand*\mcitethebibliography{\thebibliography}
\csname @ifundefined\endcsname{endmcitethebibliography}
  {\let\endmcitethebibliography\endthebibliography}{}

\end{document}